\documentclass [prb,superscriptaddress, showpacs, twocolumn]{revtex4-1}

\usepackage{graphicx}
\usepackage{dcolumn}
\usepackage{amsmath}
\usepackage{bm}
\usepackage{color}

\newcommand{\llangle}{\langle\kern-.25em\langle}
\newcommand{\rrangle}{\rangle\kern-.25em\rangle}

\makeatletter

\newcommand{\Rmnum}[1]{\expandafter\@slowromancap\romannumeral #1@}
\makeatother

\begin{document}

\title{Symmetry of re-entrant tetragonal phase in Ba$_{1-x}$Na$_x$Fe$_2$As$_2$:
Magnetic versus orbital ordering mechanism}

\author{D. D. Khalyavin}
\email{email: dmitry.khalyavin@stfc.ac.uk}
\affiliation{ISIS facility, STFC, Rutherford Appleton Laboratory, Chilton, Didcot, Oxfordshire, OX11-0QX,UK}
\author{S. W. Lovesey}
\affiliation{ISIS facility, STFC, Rutherford Appleton Laboratory, Chilton, Didcot, Oxfordshire, OX11-0QX,UK}
\affiliation{Diamond Light Source Ltd, Didcot, Oxfordshire OX11 0DE, UK}
\author{P. Manuel}
\affiliation{ISIS facility, STFC, Rutherford Appleton Laboratory, Chilton, Didcot, Oxfordshire, OX11-0QX,UK}
\author{F. Kr\"uger}
\affiliation{ISIS facility, STFC, Rutherford Appleton Laboratory, Chilton, Didcot, Oxfordshire, OX11-0QX,UK}
\affiliation{London Centre for Nanotechnology, University College London, Gordon St., London, WC1H 0AH, United Kingdom}
\author{S. Rosenkranz}
\affiliation{Materials Science Division, Argonne National Laboratory, Argonne, Illinois 60439-4845, USA}
\author{J. M. Allred}
\affiliation{Materials Science Division, Argonne National Laboratory, Argonne, Illinois 60439-4845, USA}
\author{O. Chmaissem}
\affiliation{Department of Physics, Northern Illinois University, DeKalb, Illinois 60115, USA}
\affiliation{Materials Science Division, Argonne National Laboratory, Argonne, Illinois 60439-4845, USA}
\author{R. Osborn}
\affiliation{Materials Science Division, Argonne National Laboratory, Argonne, Illinois 60439-4845, USA}

\date{\today}

\begin{abstract}
Magneto-structural phase transitions in Ba$_{1-x}$A$_x$Fe$_2$As$_2$ (A = K, Na) materials are discussed for both magnetically and orbitally driven mechanisms, using a symmetry analysis formulated within the Landau theory of phase transitions. Both mechanisms predict identical orthorhombic space-group symmetries for the nematic and magnetic phases observed over much of the phase diagram, but they predict different tetragonal space-group symmetries for the newly discovered re-entrant tetragonal phase in Ba$_{1-x}$Na$_x$Fe$_2$As$_2$  $(x \sim 0.24-0.28)$. In a magnetic scenario, magnetic order with moments along the $c$-axis, as found experimentally, does not allow any type of orbital order, but in an orbital scenario, we have determined two possible orbital patterns, specified by $P4/mnc1'$ and $I4221'$ space groups, which do not require atomic displacements relative to the parent $I4/mmm1'$ symmetry and, in consequence, are indistinguishable in conventional diffraction experiments. We demonstrate that the three possible space groups are however, distinct in resonant X-ray Bragg diffraction patterns created by Templeton $\&$ Templeton scattering. This provides an experimental method of distinguishing between magnetic and orbital models.
\end{abstract}


\maketitle

\section{Introduction}
\begin{figure}[t]
\includegraphics[scale=1.8]{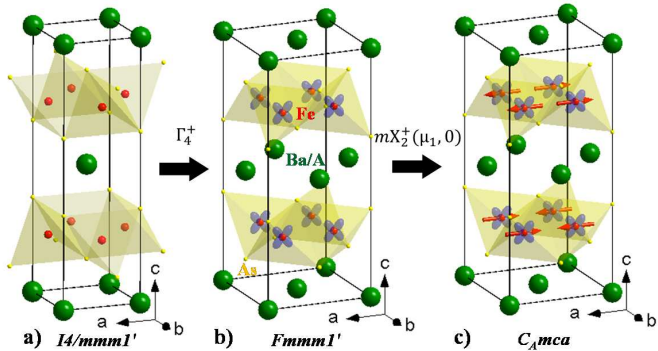}
\caption{(Color online) (a) Parent $I4/mmm1'$ crystal structure with atoms in the positions: Ba/A $2a(4/mmm1')$, Fe $4d(\bar{4}m21')$, As $4e(4mm1')$, (b) Orthorhombic $Fmmm1'$ structure of the nematic phase with atoms in the positions: Ba/A $4a(mmm1')$, Fe $8f(2221')$, As $8i(mm21')$, (c) Orthorhombic magnetic $C_Amca$ structure with atoms in the positions: Ba/A $4a(mm'm')$, Fe $8f(2'22')$, As $8g(2m'm')$.\cite{ref:6} The displayed coordinate system does not correspond to the standard setting of $C_Amca$ (see Table \ref{tab:T1} for the standard setting).}
\label{fig:F1}
\end{figure}

\indent The interplay between magnetic and structural degrees of freedom is one of the central problems in the physics of iron-based superconductors. Knowledge of the normal state from which superconductivity emerges is crucial to uncovering the true nature of the superconducting phase.\\ 
\indent In hole-doped systems, like Ba$_{1-x}$K$_x$Fe$_2$As$_2$ and Ba$_{1-x}$Na$_x$Fe$_2$As$_2$, the magnetic transition is first-order and it is associated with substantial structural distortions that reduce the symmetry from paramagnetic tetragonal $I4/mmm1'$ to magnetic orthorhombic $C_Amca (F_Cmm'm')$ [Fig. \ref{fig:F1}(a,c)]. \cite{ref:1,ref:2,ref:3,ref:4,ref:5} (We specify magnetic space groups in the Belov-Neronova-Smirnova and Opechowski-Guccione (in brackets) notations. Symmetries of phases without magnetic order are specified by gray groups. \cite{ref:6}) At first sight, it is not an unusual observation, because the magnetic order parameter $(\mu )$ is orthorhombic and therefore a coupling of the orthorhombic strain $(e_{12})$ is naturally expected through the linear-quadratic free-energy invariant, $e_{12}\mu^2$ (magneto-elastic coupling). This type of coupling implies improper critical behavior for the strain component - a critical exponent twice that for the magnetic order-parameter - in contradiction with experimental data.\cite{ref:1,ref:2,ref:3,ref:4,ref:5} Data indicate instead that the $e_{12}$ strain component is bi-linearly coupled to some other order parameter, and that the magneto-elastic contribution is small.\cite{ref:add13} This conclusion is reinforced by experimental data obtained for the electron-doped Ba(Fe$_{1-x}$Co$_x$)$_2$As$_2$ systems where magnetic and structural transitions are decoupled and are both second-order.\cite{ref:7,ref:8}  As the sample temperature decreases, the structural transition tetragonal $I4/mmm1'\rightarrow$ orthorhombic $Fmmm1'$  at $T_\textrm{nem}$ precedes the magnetic transition $Fmmm1' \rightarrow  C_Amca$  at $T_\textrm{mag}$ (Fig. \ref{fig:F1}), and the temperature gap between the two phases changes with composition.\cite{ref:7}\\
\indent The exact nature of the primary order-parameter in the nematic $Fmmm1'$ phase is undecided at the present time. Symmetry-breaking is associated with a one-dimensional time-even order parameter $(\eta )$ that transforms as the $\Gamma^+_4 ({\bm k}=0)$ irreducible representation (irrep) of the paramagnetic $I4/mmm1'$ space group. (We adopt Miller and Love notations for the special points of the $I4/mmm1'$ Brillouin zone and associated irreps as implemented into the ISOTROPY\cite{ref:10} and ISODISTORT\cite{ref:11} software used in the present study.) A specific property of $\Gamma^+_4$ is that it is not contained in the vector (mechanical) reducible representations of Ba$(2a)$, Fe$(4e)$ and As$(4d)$ Wyckoff positions. In consequence, no atomic displacive modes are allowed, which usually serve as soft modes at displacive structural phase transitions, with this symmetry. A purely ferro-elastic nature of the transition, related to the $e_{12}$ strain component as the primary order parameter (martensitic type), is very unlikely due to the similar critical temperatures in Ba$_{1-x}$A$_x$Fe$_2$As$_2$  systems with small A = Na and large A = Rb substitutional ions.\cite{ref:5,ref:12} The unit-cell volume changes in opposite ways with composition in these systems (different sign of chemical pressure which works as the driving force for proper ferro-elastic transformations\cite{ref:13}) while transition temperatures are almost the same. These observations imply that the $I4/mmm1' \rightarrow  Fmmm1'$ structural transition has a purely electronic origin. \\
\indent Two views are current about the mechanism of symmetry lowering in the nematic phase. The first one magnetic, also known as spin nematic, exploits magnetic fluctuations as the driving force for the structural distortions.\cite{ref:14,ref:15,ref:16} The magnetic order parameter $(mX^+_2)$ is two-dimensional and its components $(\mu_1,\mu_2)$ are associated with two arms of the wave vector star, ${\bm k_1}=(-1/2,1/2,0)$ and ${\bm k_2}=(1/2,1/2,0)$ ($X$-point of the $I4mmm1'$ Brillouin zone). In the widely used notations, these propagation vectors are ${\bm k_1}=(\pi,0)$ and ${\bm k_2}=(0,\pi)$ specified for the so called unfolded Brillouin zone (defined for the two-dimensional Fe-sublattice). Magnetic fluctuations between these two components become non-equivalent at $T_\textrm{nem} > T_\textrm{mag}$ which breaks the four-fold symmetry without long-range magnetic order. The second mechanism involves orbital ordering of the iron $3d$-electrons as the primary instability, which renormalizes magnetic exchange parameters in the system and triggers magnetic order.\cite{ref:17,ref:18,ref:19,ref:add3,ref:add5,ref:add6,ref:add7,ref:add8}\\
\indent Both mechanisms predict the same space-group symmetries for the nematic $Fmmm1'$ $(T_\textrm{mag} <  T < T_\textrm{nem})$ and magnetic $C_Amca$ $(T < T_\textrm{mag})$ phases, making it impossible to decide by symmetry which order parameter actually drives the transition. Very recently, a new structural transition that restores tetragonal symmetry (within the available experimental resolution) has been discovered in a Ba$_{1-x}$Na$_x$Fe$_2$As$_2$ material in a narrow range of compositions close to $x \sim 0.25$.\cite{ref:20} The re-entrant transition takes place at $T_\textrm{r}=40$-50~K, well above the critical temperature where the superconductivity emerges, $T_\textrm{c}=20$-30~K, and it is accompanied by a change in the magnetic structure.\\ 
\indent Based on theoretical predictions of an additional phase at finite doping that restores tetragonal symmetry, labeled the $C_4$ phase, and a successful refinement of the neutron diffraction data in the $I4/mmm1'$ space group (which does not remove the orbital degeneracy), the re-entrant phase transition has been interpreted as providing evidence for the magnetic mechanism.\cite{ref:20} However, the exact symmetry of the $C_4$ phase is not yet known and may provide an additional experimental method of distinguishing between magnetic and orbital ordering mechanisms. The microscopic spin-nematic calculations were based on a simplified model, in which only the iron sublattice is explicitly included, although a more complete symmetry analysis of itinerant magnetic models has been published.\cite{ref:add12} The prediction of the correct space group within the Landau theory of phase transitions\cite{ref:22} should include all atoms in the structure and all parameters that can affect the final symmetry. In particular, the analysis should include details of the magnetic structures, which were refined in the present study from neutron powder diffraction data as well as from single crystal measurements in the recent investigation reported by Wasser et al. \cite{ref:21} The main aim of the present work is, therefore, to analyze symmetry aspects of the newly-discovered, low-temperature tetragonal phase to show the symmetry-allowed space groups of the re-entrant tetragonal phase and to propose resonant x-ray experiments that may identify which is correct.\\
\begin{figure*}[t]
\includegraphics[scale=1.3]{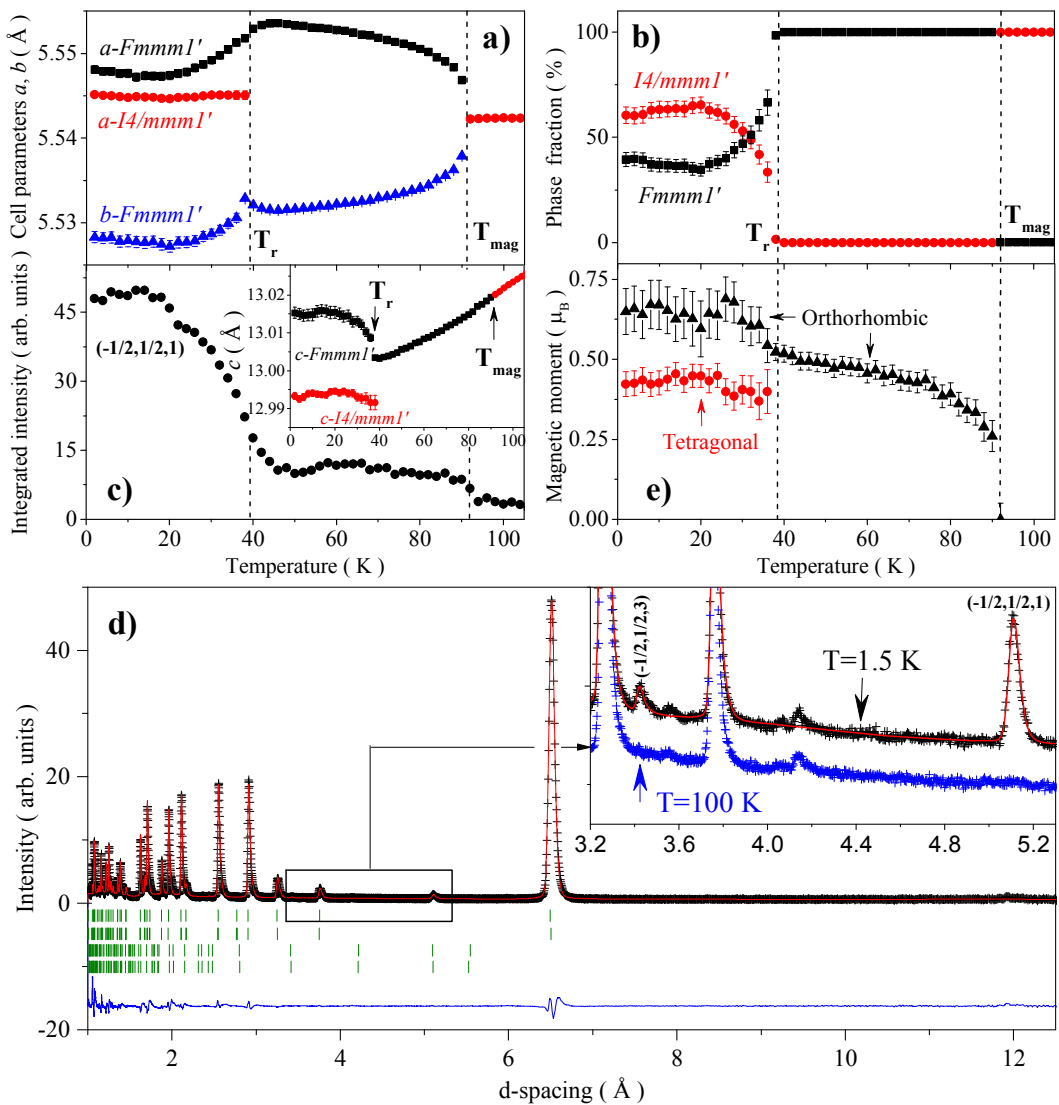}
\caption{(Color online) (a) $a$- and $b$- unit cell parameters of Ba$_{0.76}$Na$_{0.24}$Fe$_2$As$_2$ as a function of temperature (parameter for the $I4/mmm1'$ phase is multiplied by $\sqrt{2}$). (b) Phase fractions of the orthorhombic and tetragonal phases as a function of temperature. (c) Temperature dependence of integrated intensity of the magnetic $(-1/2,1/2,1)$ reflection. Inset shows the temperature dependence of the unit cell parameter $c$. (d) Rietveld refinement of the neutron powder diffraction pattern collected on the WISH diffractometer (ISIS). The cross symbols (black) and solid line (red) represent the experimental and calculated intensities, respectively, and the line below (blue) is the difference between them. Tick marks indicate the positions of Bragg peaks. The first two rows from the top correspond to the nuclear $I4/mmm1'$ and $Fmmm1'$ phases, the second two rows represent their magnetic counterparts. Insert shows the patterns collected at 1.5 K and 100 K, at a vicinity of the strongest magnetic peaks. (e) Ordered moment as a function of temperature in the orthorhombic and re-entrant tetragonal phases.}
\label{fig:F1_1}
\end{figure*}
\indent The paper is organized as follows. In Section II, we provide experimental neutron diffraction data collected for the Ba$_{0.76}$Na$_{0.24}$Fe$_2$As$_2$ composition exhibiting evidence of the re-entrant tetragonal phase. In Section III, we analyze the magnetic and orbital ordering mechanisms for symmetry lowering in both the orthorhombic and tetragonal phases. In the magnetic scenario, the assumption that the symmetry of the re-entrant phase is fully determined by the magnetic order parameter sets constraints on the possible orbital patterns (if any) which are compatible with it. In the orbital scenario, the symmetry is determined by the intersection between the symmetry of the orbital pattern and the triggered magnetic order parameter. The corresponding orbital order in the re-entrant phase should not allow coupling to any atomic displacements and symmetry-breaking strain components. Instead, the local symmetry of sites used by Fe ions should break the orbital degeneracy. Compatible orbital ordered patterns have been predicted based on microscopic spin-orbital model.\cite{ref:17} Allowed isotropy subgroups are shown in Section III.C to be $P4/mnc1'$ and $I4221'$, which both keep the original setting and origin of the parent group. By construction, the subgroups possess identical extinction rules for Bragg diffraction of neutrons and X-rays. We demonstrate in Section IV that the technique of resonant X-ray Bragg diffraction can distinguish between $P4/mnc1'$ and $I4221'$ type structures that would result from an orbitally driven scenario and the $I4/mmm1'$ space group
that is predicted for the magnetically driven mechanism. Previously, resonant X-ray Bragg diffraction has been used to confirm a similar purely electronic transition in neptunium dioxide (NpO$_2$). In this case, the reduction of the fluorite structure $Fm\bar{3}m1'$ to $Pn\bar{3}m1'$ also does not involve atomic displacements.\cite{ref:22a} Our simulation of resonant x-ray diffraction in Section IV is not unlike that reported for $Fm\bar{3}m1' \rightarrow Pn\bar{3}m1'$.\cite{ref:22b} Conclusions from our work are found in Section V.\\

\section{Experimental evidence of re-entrant phase in B\lowercase{a}$_{0.76}$N\lowercase{a}$_{0.24}$F\lowercase{e}$_2$A\lowercase{s}$_2$}
The high-resolution neutron powder diffraction data collected for the Ba$_{0.76}$Na$_{0.24}$Fe$_2$As$_2$ composition above 90 K were successfully refined in the tetragonal $I4/mmm1'$ space group (for details of the neutron diffraction experiment see Ref.[\onlinecite{ref:20}]). At $T_\textrm{mag}$=90 K, the first order phase transition to the magnetic orthorhombic $C_Amca$ phase is evidenced by splitting some of fundamental peaks and appearing additional Bragg reflections consistent with the propagation vector ${\bm k}=1/2,1/2,0$. This orthorhombic phase involves stripe-type antiferromagnetic ordering which is typical for the potassium and sodium-doped compositions with $x<0.24$. Below $T_\textrm{r}$=40 K, the transition to the re-entrant tetragonal phase takes place as reported in Ref.[\onlinecite{ref:20}]. The transition is not complete and the re-entrant phase coexists with the orthorhombic one down to the lowest measured temperature 1.5 K. Inspection of the diffraction patterns and the qualitative Rietveld refinement did not reveal any evidence of the symmetry lowering in the nuclear structure of the re-entrant phase in comparison with the structure of the high-temperature paramagnetic phase and therefore, the nuclear scattering for the re-entrant phase was modeled using the parent $I4/mmm1'$ symmetry. The scattering for the orthorhombic phase was done in the $Fmmm1'$ space groups. The unit cell parameters and the phase fractions as a function of temperature are shown in Figure \ref{fig:F1_1}a and \ref{fig:F1_1}b, respectively. The unit cell of the re-entrant tetragonal phase is stretched in the $(ab)$ plane and shrunken along the $c$-axis compared to the paramagnetic one. The refinement indicates that the transition at $T_\textrm{r}$ results in changing of the unit cell parameters of the orthorhombic phase as well. The coupling between the phases can be caused by the internal strains appearing on the phase boundaries and therefore the coupling strength might depend on the microstructure and can vary from one sample to another.\\
\indent The transition to the re-entrant phase involves also modification of the magnetic scattering as indicated by the temperature dependence of the integrated intensity of the $(-1/2,1/2,1)$ magnetic reflection (Fig. \ref{fig:F1_1}c). The propagation vector of the magnetic structure does not change across the transition but the structure of the low-temperature tetragonal phase is different from the higher-temperature orthorhombic one. Assuming irreducible nature of the magnetic order parameter in the tetragonal phase, the best refinement quality ($R_\textrm{magnetic}=7.47 \%$)  of the powder diffraction data (Fig. \ref{fig:F1_1}d) was obtained in the antiferromagnetic stripe model with the magnetic dipoles polarized along the $c$-axis, in agreement with the recent single crystal study of Wasser et al.\cite{ref:21} Thus, the main impact of the transition at $T_\textrm{r}$ on the magnetic structure of Ba$_{0.76}$Na$_{0.24}$Fe$_2$As$_2$ is swapping the moments direction from in-plane to out-of-plane in the re-entrant phase. The magnetic structure of the low-temperature orthorhombic phase (coexisting with the tetragonal one) was found to be qualitatively identical to the structure of the higher-temperature phase. The temperature dependence of the magnetic moments refined independently for the orthorhombic and tetragonal phases is shown in Fig. \ref{fig:F1_1}e. The ordered moment of the tetragonal phase is notably smaller than the moment of the orthorhombic phase assuming single-${\bm k}$ magnetic structures. The symmetry aspects of these models as well as their two-${\bm k}$ counterparts (which are indistinguishable in the powder diffraction data) are discussed in the next section.\\
\begin{figure}[t]
\includegraphics[scale=2.15]{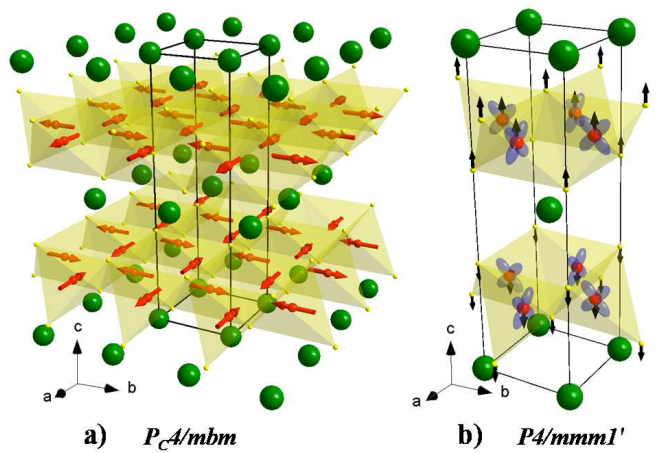}
\caption{(Color online) (a) Tetragonal magnetic structure with the $P_C4/mbm$ space group, involving two propagation vectors ${\bm k_1}=(-1/2,1/2,0)$ and ${\bm k_2}=(1/2,1/2,0)$ and atoms in the positions: Ba/A1 $2a(4'/mm'm)$, Ba/A2 $2b(4'/mm'm)$, Fe $8i(2.mm)$, As1 $4g(4'm'm)$, As2 $4h(4'm'm)$.\cite{ref:6} Only the unit cell of the parent $I4/mmm1'$ structure is displayed (see Table \ref{tab:T1} for the choice of the magnetic cell). (b) Atomic displacements (and orbital ordering) with $P4/mmm1'$ symmetry, allowed in the magnetic  $P_C4/mbm$ space group as a secondary order parameter, which contribute to the $h+k+l$ odd reflections with $l\neq0$.}
\label{fig:F2}
\end{figure}
\section{Analysis of mechanisms for symmetry lowering}
\subsection{Magnetic mechanism with in-plane moments}
\indent We start with the symmetry analysis of the re-entrant tetragonal phase of Ba$_{1-x}$Na$_x$Fe$_2$As$_2$ with magnetic moments in the $(ab)$-plane. The published spin-nematic calculations did not include spin-orbit coupling so no moment direction was defined, but this is one of the possible ground states discussed within a magnetic scenario.\cite{ref:14,ref:15,ref:16,ref:add12} It seems to be in contradiction with the neutron diffraction data presented in the previous Section and recently reported by Wasser et al.\cite{ref:21}, which both indicate that the moments are parallel with the $c$-axis, but we include this discussion for completeness and uniform consideration of some symmetry aspects of the transitions at $T_\textrm{mag}$ and $T_\textrm{r}$ as well for possible relevance to other systems.\\
 \begin{table*}[t]
\caption{Equilibrium order parameter directions in the $mX^+_2$ representation space and the magnetic space groups for the four stable phases obtained by minimization of the free-energy (\ref{eq:E1}). Columns "basis" and "origin" represent the basis vectors and the origin choice of the magnetic subgroups, respectively, in respect of the parent $I4/mmm1'$ space group. The magnetic space groups for the case of the reducible $\Gamma^+_4 \oplus mX^+_2$ order parameter are given as well.}
\centering 
\begin{tabular*}{0.98\textwidth}{@{\extracolsep{\fill}} l l l l r} 
\\
\hline\hline\\  [-1.5ex]
Irrep & Order parameter & Space group & Basis & Origin \\ [1.0ex] 
\hline\\[-1.5ex] 
$mX^+_2$ & $\mu_1=\mu_2=0$ & $I4/mmm1'$ & $(1,0,0)(0,1,0)(0,0,1)$ & $(0,0,0)$\\ 
$mX^+_2$ & $\mu_1 \neq0,\mu_2=0$ & $C_Amca (F_Cmm'm')$ & $(0,0,1)(1,1,0)(-1,1,0)$ & $(0,0,0)$\\
$mX^+_2$ & $\mu_1=\mu_2 \neq 0$ & $P_C4/mbm (P_P4'/mmm')$ & $(-1,1,0)(-1,-1,0)(0,0,1)$ & $(-1/2,1/2,0)$\\
$mX^+_2$ & $\mu_1 \neq 0,\mu_2 \neq 0,\mu_1 \neq \mu_2$ & $P_Cbam (C_Pm'm'm)$ & $(-1,1,0)(-1,-1,0)(0,0,1)$ & $(0,0,0)$ \\ [1.0ex]
\hline\\ [-1.5ex]
$\Gamma^+_4 \oplus mX^+_2$ & $\eta \neq 0, \mu_1=0,\mu_2=0$ & $Fmmm1'$ & $(1,1,0)(-1,1,0)(0,0,1)$ & $(0,0,0)$ \\
$\Gamma^+_4 \oplus mX^+_2$ & $\eta \neq 0, \mu_1 \neq0,\mu_2=0$ & $C_Amca (F_Cmm'm')$ & $(0,0,1)(1,1,0)(-1,1,0)$ & $(0,0,0)$ \\
$\Gamma^+_4 \oplus mX^+_2$ & $\eta \neq 0, \mu_1 \neq 0,\mu_2 \neq 0$\footnotemark[1] & $P_Cbam (C_Pm'm'm)$ & $(-1,1,0)(-1,-1,0)(0,0,1)$ & $(0,0,0)$ \\ [1.0ex]
\hline\hline  
\end{tabular*}
\footnotetext[1]{the cases when $\mu_1 \neq \mu_2$ and $\mu_1=\mu_2$ both result in the same orthorhombic $P_Cbam$ symmetry at $\eta \neq 0$}
\label{tab:T1} 
\end{table*} 
\indent When the magnetic moments are confined within the $(ab)$-plane, as experimentally found in the orthorhombic phase of all Ba$_{1-x}$A$_x$Fe$_2$As$_2$ pnictides, the magnetic order-parameter is associated with the time-odd and two-dimensional irrep $mX^+_2$ of the parent $I4mmm1'$ space group.\cite{ref:23} The two components of the order parameter are related to the ${\bm k_1}=(-1/2,1/2,0)$ and ${\bm k_2}=(1/2,1/2,0)$ propagation vectors of the $I4mmm1'$ Brillouin zone. The integrity basis consists of the two polynomial invariants $\mu_1^2+\mu_2^2$ and $\mu_1^2 \mu_2^2$, which results in the Landau free-energy decomposition:
\begin{align}
&F(\mu_1,\mu_2) = a_1(\mu_1^2 + \mu_2^2) + b_1(\mu_1^4 + \mu_2^4) + \nonumber \\ 
&b_2(\mu_1^2  \mu_2^2) + c_1(\mu_1^6 + \mu_2^6) + c_2(\mu_1^4 \mu_2^2 + \mu_1^2 \mu_2^4) + \nonumber \\
&d_1(\mu_1^8 + \mu_2^8) + c_2(\mu_1^6 \mu_2^2 + \mu_1^2 \mu_2^6) + d_3(\mu_1^4  \mu_2^4) + \cdot \cdot \cdot 
\label{eq:E1}
\end{align}
Minimization of $F(\mu_1,\mu_2)$ truncated at the eighth degree leads to four possible stable phases.\cite{ref:24,ref:25} The equilibrium order parameters and the
 magnetic symmetries (for the case of the $mX^+_2$ representation) of the corresponding phases are listed in Table \ref{tab:T1}. The symmetry of the phase where both components of the magnetic order parameter are non-zero and equal $(\mu_1=\mu_2 \neq 0)$ is tetragonal $P_C4/mbm$. \\
\indent The magnetic $P_C4/mbm$ structure involves two propagation vectors and implies the orthogonal spin configuration shown in Fig. \ref{fig:F2}(a). A structural distortion $(\xi )$ violating the body centring condition - Miller indexes with $h + k + l$ even - is expected in the case of the $P_C4/mbm$ symmetry as a secondary order-parameter through the magneto-elastic coupling $\xi \mu_1 \mu_2 \equiv \xi \mu^2$ for $\mu_1=\mu _2=\mu$. Corresponding distortive modes are associated with the $M^+_1$ $[{\bm k} = (1,1,1)]$ irrep of $I4/mmm1'$ and involve displacements of both Fe and As ions along the $c$-axis, lowering the symmetry down to $P4/mmm1'$ [Fig. \ref{fig:F2}(b)]. These displacements allow Bragg reflections $h + k + l$ odd with $l \neq 0$, which can be observed in conventional diffraction experiments using X-rays or neutrons. No Bragg reflections of this type were visible in neutron powder diffraction measurements.\cite{ref:20} It is interesting to note that the $P_C4/mbm$ space group allows the orbital ordering shown in Fig. \ref{fig:F2}(b) since the site symmetry of the Fe ions is $2.mm$. This orbital pattern with the $M^+_1$ symmetry is coupled to the primary magnetic order parameter $mX^+_2$ through the magnetoelastic coupling invariant specified above.\\
\indent The free-energy given by Eq. \ref{eq:E1} is only a `minimal' decomposition necessary to discuss the symmetry of the re-entrant phase using a single irreducible magnetic order parameter. To take into account the transition to the nematic $Fmmm1'$ phase at $T_\textrm{nem}$, an additional time-even order parameter must be included into the decomposition. The symmetry breaking at the $I4/mmm1' \rightarrow  Fmmm1'$ transition is associated with the one-dimensional irreducible representation $\Gamma^+_4$. The primary order parameter, $\eta $, has the symmetry of this representation and is linearly coupled to the $e_{12}$ strain component. For our symmetry discussion, the explicit physical meaning of $\eta $ is not essential (it can be either Ising spin nematic or orbital ordering order parameter).\\
\begin{table*}[t]
\caption{Equilibrium order parameter directions in the $mX^+_3$ representation space and the magnetic space groups for the four stable phases obtained by minimization of the free-energy (\ref{eq:E1}). Columns ``basis" and ``origin" represent the basis vectors and the origin choice of the magnetic subgroups in respect of the parent $I4/mmm1'$ space group.}
\centering 
\begin{tabular*}{0.98\textwidth}{@{\extracolsep{\fill}} l l l l r} 
\\
\hline\hline\\  [-1.5ex]
Irrep & Order parameter & Space group & Basis & Origin \\ [1.0ex] 
\hline\\[-1.5ex] 
$mX^+_3$ & $\mu_3=\mu_4=0$ & $I4/mmm1'$ & $(1,0,0)(0,1,0)(0,0,1)$ & $(0,0,0)$\\ 
$mX^+_3$ & $\mu_3 \neq0,\mu_4=0$ & $C_Amca (F_Cmm'm')$ & $(-1,-1,0)(0,0,-1)(1,-1,0)$ & $(0,0,0)$\\
$mX^+_3$ & $\mu_3=\mu_4 \neq 0$ & $P_C4_2/ncm (P_P4_2/m'mc)$ & $(-1,1,0)(-1,-1,0)(0,0,1)$ & $(1/2,-1/2,-1/2)$\\
$mX^+_3$ & $\mu_3 \neq 0,\mu_4 \neq 0,\mu_3 \neq \mu_4$ & $P_Cccn (C_Pccm')$ & $(-1,1,0)(-1,-1,0)(0,0,1)$ & $(0,0,0)$ \\ [1.0ex]
\hline\hline  
\end{tabular*}
\label{tab:T2} 
\end{table*}
\indent The extended free-energy containing coupling terms $\eta (\mu_1^2 - \mu_2 ^2)$ and $\eta^2 (\mu_1^2 + \mu_2 ^2)$ describes three additional phases with the symmetries specified in Table \ref{tab:T1} for the case of the coupled $\Gamma^+_4 \oplus mX^+_2$ order parameter. Below $T_\textrm{mag}$, the long-range magnetic ordering associated with the $mX^+_2$ representation and the $(\mu_1, 0)$ order parameter direction usually takes place as a second-order phase transition from the parent (for this transition) symmetry $Fmmm1'$. The continuous nature of this transition implies that the magnetic phase must be the result of a common action of the two order parameters, time-even $\Gamma^+_4(\eta)$ and time-odd $mX^+_2(\mu_1,0)$. The symmetry of this reducible order-parameter $\Gamma^+_4(\eta ) \oplus  mX^+_2(\mu_1,0)$ is $C_Amca$ (see Table \ref{tab:T1}) so it is identical to the symmetry of the $mX^+_2(\mu_1,0)$ order parameter alone.\\
\indent In spite of the identical symmetry, the cases of the reducible and irreducible order-parameters are essentially different and this is the key point at this stage. For instance, a transition from the phase with the reducible order-parameter to the phase with the irreducible one implies renormalization of the coupling for the $e_{12}$ strain component to linear-quadratic and must be necessarily first order (as any isostructural transition). Another crucial point is that in the case of the reducible order-parameter $\Gamma^+_4 \oplus  mX^+_2$, a condensation of the second component of the magnetic order parameter $mX^+_2(\mu_1 = \mu_2 \neq 0)$ will not restore tetragonal symmetry. The resultant symmetry will still be an intersection between $\Gamma^+_4(\eta )$ and $mX^+_2(\mu_1 = \mu_2 \neq 0)$ which results in the orthorhombic $P_Cbam$ magnetic space group (Table \ref{tab:T1}).\\
\indent This is the fundamental difference between reducible and irreducible order-parameters; only in the latter case can a condensation of additional components increase the symmetry of the system. Thus, the continuous nature of the $Fmmm1' \rightarrow  C_Amca$ transition (resulting in the reducible order-parameter) and the crossover to the tetragonal $P_C4/mbm$ phase (requiring an irreducible order-parameter) are mutually exclusive, if one assumes the purely magnetic nature of the transition to the re-entrant phase. The low temperature phase transition must necessarily involve a structural (electronic) instability which cancels the $\Gamma^+_4(\eta)$ time-even order parameter. In the case of the orbital reconstruction mechanism, $\eta $ is replaced by another order parameter and the symmetry of the system is determined by the intersection between the symmetry of the new orbital pattern and the triggered magnetic order parameter.\\
Note that, a similar conclusion about the reducible character of the distortions is applicable for the case of the single magneto-structural $I4/mmm1' \rightarrow  C_Amca$ phase transition with identical critical behavior for the orthorhombic strain component $e_{12}$ and the magnetic order-parameter as experimentally observed in the Ba$_{1-x}$K$_x$Fe$_2$As$_2$ and Ba$_{1-x}$Na$_x$Fe$_2$As$_2$ systems.\cite{ref:3,ref:4,ref:5} This critical behavior indicates that the $e_{12}$ strain component is not induced by the magnetic order-parameter as a secondary distortion through the magneto-elastic coupling. Instead, this behavior points to a linear coupling between $e_{12}$ and some other order parameter, $\eta $, having its own instability near the transition temperature. The bi-quadratic relation between $\eta $ and the magnetic order-parameter implies a reducible nature of the distortions in the $C_Amca$ phase and indicates that the coupling between these order parameters has a microscopic origin rather than symmetry-related one.\\
\subsection{Magnetic mechanism with out-of-plane moments}
\begin{figure}[b]
\includegraphics[scale=2.0]{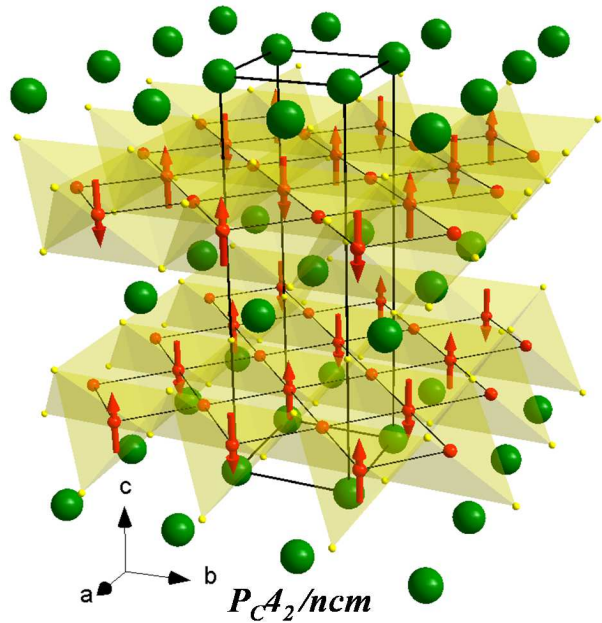}
\caption{(Color online) Tetragonal magnetic structure with the $P_C4_2/ncm$ space group, involving two propagation vectors ${\bm k_1}=(-1/2,1/2,0)$ and ${\bm k_2}=(1/2,1/2,0)$ and atoms in the positions: Ba/A $4c(m'.mm')$, Fe1 $4e(\bar{4}'2'm)$ - the site with zero magnetic dipole moment, Fe2 $4f(\bar{4}2'm')$ - the site with non-zero magnetic dipole moment, As $8i(2.mm)$.\cite{ref:6} Only the unit cell of the parent $I4/mmm1'$ structure is displayed (see Table \ref{tab:T2} for the choice of the magnetic cell).}
\label{fig:F2a}
\end{figure}
\indent The tetragonal $P_C4/mbm$ space group obtained in the previous section with the irreducible $mX^+_2$ magnetic order parameter seems to be irrelevant to the case of the re-entrant phase in Ba$_{1-x}$Na$_x$Fe$_2$As$_2$, since it has been experimentally shown that the magnetic moments are along the $c$-axis (see Section II and Ref.[\onlinecite{ref:21}]). Thus, to adopt the magnetic mechanism for these experimental findings, we have to introduce in our phenomenological approach another magnetic order parameter $(\mu_3,\mu_4)$ with the symmetry of the $mX^+_3$ irrep which transforms the out-of-plane components of the magnetic dipoles with $\bm {k_1}=(-1/2,1/2,0)$  and $\bm {k_2}=(1/2,1/2,0)$ propagation vectors. Note that this is not forbidden by symmetry since the transition to the re-entrant phase is strongly first order. The image group of $mX^+_3$ and therefore the free-energy decomposition is identical to the previous case of the $mX^+_2$ irrep. The equilibrium order parameters obtained by minimization of the functional (\ref{eq:E1}) 
correspond to the stable magnetic phases for $mX^+_3$ listed in Table \ref{tab:T2}. The tetragonal space group $P_C4_2/ncm$ with $\mu_3=\mu_4\neq0$ is the symmetry of the system in the adopted magnetic scenario. It should be pointed out that the proper phenomenological approach for the magnetic mechanism, which describes both $C_Amca$ $(\mu_1 \neq0,\mu_2=0)$ and $P_C4_2/ncm $ $(\mu_3=\mu_4\neq0)$ magnetic phases should be based on the Landau decomposition written in terms of the reducible $mX^+_2 \oplus  mX^+_3$ order parameter components, $\mu_1,\mu_2,\mu_3,\mu_4$.  Minimization of this functional yields six more 'mixed' phases where some of the components of both $mX^+_2$ and $mX^+_3$ order parameters are non-zero. The corresponding magnetic structures combine both in-plane and out-of-plane configurations, but since there are no solutions with tetragonal symmetry between the 'mixed' phases, we do not consider them any further.\\
\begin{figure}[t]
\includegraphics[scale=1.8]{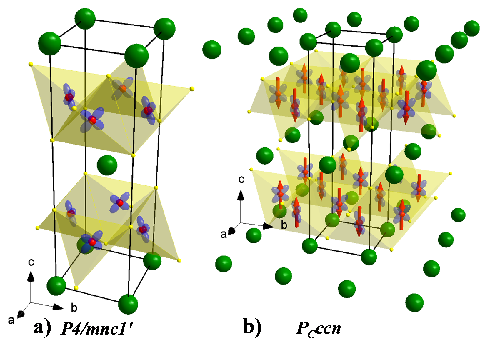}
\caption{(Color online) (a) Orbital ordering associated with the $M^+_3$ irrep resulting in the tetragonal $P4/mnc1'$ space group. The crystallographic positions occupied by atoms in $P4/mnc1'$ are: Ba/A $2a(4/m..1')$, Fe $4d(2.221')$, As $4e(4..1')$ (b) Combination of the $M^+_3$ orbital ordering with the $mX^+_3 (\mu_3 \neq 0, \mu_4=0)$ magnetic order, resulting in the $P_Cccn$ magnetic space group with atoms in the positions: Ba/A $4e(..2'/m')$, Fe1 $4a(2'2'2)$, Fe2 $4b(2'2'2)$, As $8k(..2')$.\cite{ref:6}}
\label{fig:F3}
\end{figure}
\indent The two-${\bm k}$ magnetic structure with tetragonal $P_C4_2/ncm$ symmetry imposes zero dipole magnetic moments for half of the Fe sites  (Fig. \ref{fig:F2a}). The remarkable feature is that the site symmetry of Fe in the $4e$ and $4f$ Wyckoff positions with zero and non-zero magnetic dipole moments are $(\bar{4}'2'm)$ and $(\bar{4}2'm')$, respectively. These site symmetries do not remove the degeneracy between the $d_{xz}$ and $d_{yz}$ orbitals and their linear combinations. In other words, the symmetry does not permit any type of orbital ordering and therefore the 'chicken and egg' question, whether magnetism drives orbital ordering or vice versa (the major issue in the orthorhombic magnetic phase), does not exist for this phase. It can be driven only by magnetic instability.
Thus, proving experimentally the two-$\bm{k}$ nature of the magnetic order in the re-entrant phase, for instance by neutron diffraction experiment with uniaxial strain applied to the crystal, would provide strong evidence for the magnetically driven scenario. \\
\indent Contrary to the case with the in-plane moments, the out-of-plane tetragonal magnetic structure does not permit any atomic displacements and keeps all the atoms in the same positions as they are in the parent $I4/mmm1'$ space group. The magnetic order parameter allows a magnetoelastic coupling invariant with time-odd physical quantities $(\xi)$ transforming as $M^+_2$ irrep, $\xi \mu_3 \mu_4 \equiv \xi \mu^2$ for $\mu_3=\mu _4=\mu$. This coupling, however, does not change the site symmetry of Fe and therefore in the diffraction experiment, discussed in the section IV, the crystal structure symmetry of the system can be well approximated by the parent $I4/mmm1'$ space group.
\begin{figure}[t]
\includegraphics[scale=1.8]{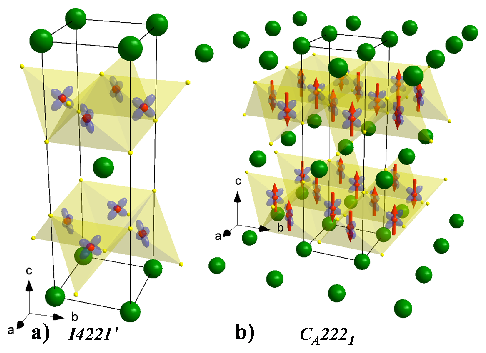}
\caption{(Color online) (a) Orbital ordering associated with the $\Gamma^-_1$ irrep resulting in the tetragonal $I4221'$ space group. The crystallographic positions occupied by atoms in $I4221'$ are: Ba/A $2a(4221')$, Fe $4d(2.221')$, As $4e(4..1')$ (b) Combination of the $\Gamma^-_1$ orbital ordering with the $mX^+_3 (\mu_3 \neq 0, \mu_4=0)$ magnetic order, resulting in the $C_A222_1$ magnetic space group with atoms in the positions: Ba/A $4a(22'2')$, Fe1 $4c(2'22')$, Fe2 $4d(2'22')$, As $8f(.2'.)$.\cite{ref:6}}
\label{fig:F4}
\end{figure}
\subsection{Orbital ordering mechanism}
\begin{table*}[t]
\caption{The magnetic space groups for the case of the $M^+_3\oplus mX^+_3$ and $\Gamma^-_1\oplus mX^+_3$ reducible order parameters.}
\centering 
\begin{tabular*}{0.98\textwidth}{@{\extracolsep{\fill}} l l l l r} 
\\
\hline\hline\\  [-1.5ex]
Irrep & Order parameter & Space group & Basis & Origin \\ [1.0ex] 
\hline\\ [-1.5ex]
$M^+_3(\eta ) \oplus mX^+_3 (\mu_3,\mu_4)$ & $\eta \neq 0, \mu_3 \neq 0,\mu_4=0$ & $P_Cccn (C_Pccm')$ & $(1,-1,0)(1,1,0)(0,0,1)$ & $(0,0,0)$ \\
$\Gamma^-_1(\eta ) \oplus mX^+_3(\mu_3,\mu_4)$ & $\eta \neq 0, \mu_3 \neq 0,\mu_4=0$ & $C_A222_1 (F_C22'2')$ & $(1,1,0)(0,0,1)(1,-1,0)$ & $(1/4,1/4,0)$ \\ [1.0ex]
\hline\hline  
\end{tabular*}
\label{tab:T3} 
\end{table*}
\indent In this mechanism the primary instability is related to a spontaneous change of the $d_{xz}$ and $d_{yz}$ orbital occupancies that reduce Fe-site symmetry from $\bar{4}m21'$ to $2221'$. Site symmetry breaking is associated with the $B_1$ point group representation, subduced by the $\Gamma^+_4$ space group irrep that induces global orthorhombic distortions.\cite{ref:10,ref:11} The macroscopic strain component $e_{12}$ transforms as the $\Gamma^+_4$ irrep as well, resulting in a linear coupling to the primary order parameter, $e_{12}\eta $. In this scenario, an electronic instability renormalizes exchange parameters in the system and triggers a magnetic ordering, thus the symmetry of the system, is always an intersection between the orbital ordering and magnetic order parameters.\\
\indent It should be pointed out that the interaction between orbital ordering and magnetic order parameters is caused by microscopic reasons rather than symmetry. This means that the dominant phenomenological free-energy coupling term should not be necessarily the lowest degree one, as for the case of secondary order-parameters, and depends on the explicit form of this interaction. Experimental data\cite{ref:3,ref:4,ref:5} indicate the dominant role of the quadratic-quadratic free-energy invariant $\eta^2\mu^2$ representing the linear part of the interaction.\\
\indent Since $\Gamma^+_4$ is a one-dimensional order parameter, the orbital reconstruction in the re-entrant tetragonal phase of Ba$_{1-x}$Na$_x$Fe$_2$As$_2$ must be associated with another irreducible representation. The high resolution neutron diffraction data (nuclear structure) were successfully refined in the parent $I4/mmm1'$ space group (see Section II and Ref.[\onlinecite{ref:20}]). This symmetry does not remove the orbital degeneracy and, therefore, in the orbital-ordering mechanism the actual symmetry must be different. To be consistent with the experimental data, we should assume that the orbital ordering in the re-entrant phase does not allow coupling to any atomic displacements and symmetry-breaking strain components. In addition, the site symmetry of the crystallographic position used by Fe should break the orbital degeneracy. These symmetry conditions can be reformulated in an exact group-theoretical way and rigorously checked.\\ 
\indent The desired isotropy subgroup should be associated with a space group irreducible representation which is induced by the Fe-site irrep $B_1$, whose subduction frequency is zero for all reducible vector representations in the structure. The relevant analysis reveals that only two one-dimensional irreducible representations of the $I4/mmm1'$ space group satisfy both conditions, namely, $M^+_3 [{\bm k} = (1, 1, 1)]$ and $\Gamma^-_1 ({\bm k} = 0)$.\cite{ref:10,ref:11} The corresponding isotropy subgroups are $P4/mnc1'$ and $I4221'$ which both keep the original setting and origin of the parent group. Symmetry reductions $I4/mmm1' \rightarrow  P4/mnc1'$ and $I4/mmm1' \rightarrow I4221'$ are caused by losing site symmetry alone, with no atomic displacements relative to the parent $I4/mmm1'$ structure. This means that the conventional Rietveld analysis of neutron or conventional X-ray diffraction data is not able to reveal the actual structural symmetry. The orbital patterns associated with the $P4/mnc1'$ and $I4221'$ subgroups are shown in Figs. \ref{fig:F3}(a) and \ref{fig:F4}(a), respectively. They represent an alternation of the $d_{xz}$ and $d_{yz}$ orbitals in the $(ab)$ plane and are different in the way of stacking the ordered layers along the $c$-axis. In fact, if one considers only the two-dimensional layers formed by Fe-ions, these patterns are identical to the antiferro $O(\pi,\pi)$ orbital state in the original work by Kr\"uger \emph{et al} Ref.[\onlinecite{ref:17}]. This type of orbital ordering is stable in a wide parametric space (see the phase diagram in Fig. 4 of Ref.[\onlinecite{ref:17}]) and has a common phase boundary with the ferrorbital $O(0,0)$ orthorhombic phase. Moreover, examination of the structural parameter, $\lambda $, controlling the stability of the orbitally ordered phases, as a function of Na-doping in Ba$_{1-x}$Na$_x$Fe$_2$As$_2$, indicates that the system moves in the right direction towards the $O(0,0)\rightarrow O(\pi,\pi)$ transition.\\
\indent The lifting of the orbital degeneracy in the $P4/mnc1'$ and $I4221'$ structures would be a purely electronic effect without any structural signature (no structural distortions are allowed apart from the non-symmetry breaking strain component $e_{33}$). If confirmed, it would represent a very unusual situation in comparison with other known orbitally-ordered systems (like manganites or cuprates) where the orbital and lattice degrees of freedom are intimately related and lifting of orbital degeneracy is manifested by local distortions of the coordinated structural units.\\
\indent There are no symmetry restrictions on the magnetic order-parameter and the new orbital pattern may trigger different magnetic configurations. One of the probable candidates for the magnetic structure in the re-entrant phase, which provides a good fit to magnetic intensities (see Fig.\ref{fig:F1_1}d (inset) and Ref.[\onlinecite{ref:21}]) implies a propagation vector ${\bm k} = (1/2, 1/2, 0)$ and magnetic dipole moments polarized along the $c$-axis. This magnetic configuration is associated with the $mX^+_3$ irrep as specified in the previous section. The stability of this configuration in terms of the nearest and next nearest neighbor exchange interactions has been discussed by Kr\"uger \emph{et al} in Ref.[\onlinecite{ref:17}]. Between the equilibrium phases listed in Table \ref{tab:T2}, only the magnetic configuration with $C_Amca$ symmetry keeps magnetic moments constant on all the Fe-sites. Note, that the space group symbol is identical to the phase with the same order parameter direction in the $mX^+_2$ irrep from Table \ref{tab:T1}, the unit cell choice is however different in both cases, which implies different magnetic structures. The tetragonal phase $P_C4_2/ncm$ imposes zero ordered moment for half of the sites. Although this can occur in an itinerant magnetic scenario,\cite{ref:add12} it is unlikely in an orbital scenario with localized electrons because of the large entropy that it entails.\\
\indent A combination of the $mX^+_3 (\mu_3,0)$ magnetic order-parameter with the orbital ordering having $M^+_3$ and $\Gamma^-_1$ symmetries results in $P_Cccn (C_Pccm')$ [Fig. \ref{fig:F3}(b)] and $C_A222_1 (F_C22'2')$ [Fig. \ref{fig:F4}(b)] magnetic space groups, respectively (Table \ref{tab:T3}). In both cases, the resultant magnetic symmetry is orthorhombic which does not allow atomic displacements relative to a tetragonal $I4/mmm1'$ structure, but it permits a coupling to the symmetry breaking strain component $e_{12}$. In the powder neutron diffraction experiment (see Section II and Ref.[\onlinecite{ref:20}]), this component was not detected but this is possible if the magneto-elastic coupling is weak.\\
\indent The key point is that the orbital patterns with the $P4/mnc1'$ and $I4221'$ symmetries cannot be induced by any magnetic order parameter associated with the $mX^+_3$ or $mX^+_2$ irreps.\cite{footnote} The magnetic space groups listed in Tables \ref{tab:T1} and \ref{tab:T2} forbid this kind of orbital ordering and therefore the $P4/mnc1'$ and $I4221'$ patterns can appear only as a result of the electronic instability unrelated to the magnetic degree of freedom. Therefore, an experimental observation of one of these patterns in the X-ray resonant experiment discussed in the next section would be unambiguous evidence for an orbitally driven mechanism.
\section{X-ray resonant scattering}
\indent We calculate unit-cell structure factors for Bragg diffraction by $P4/mnc1'$ and $I4221'$ type structures, labelled (A) and (C), to unveil signatures of the orbital ordering. A calculation for the $I4/mmm1'$ type structure, labelled (B), provides a reference point to our findings.\\
\indent Structure factors for Templeton $\&$ Templeton (T $\&$ T) scattering are made functions of the angle of rotation of a crystal about the Bragg wave vector - an azimuthal-angle scan.\cite{ref:26,ref:27} Bulk properties of a material, subject to elements of symmetry in the crystal class, are revealed in a structure factor evaluated for Miller indices $h = k = l = 0$, i.e., the forward direction. Intensities of non-trivial Bragg spots $(h, k, l)$ depend on translations in the unit cell and the symmetry of sites used by resonant ions. Our calculations include each and everyone of the elements of symmetry in a space group. This is conveniently achieved with a theory of resonant scattering that uses atomic multipoles, defined to possess discrete symmetries with respect to inversion of space coordinates and the reversal of the direction of time.\cite{ref:28,ref:29,ref:30} In the present work we discuss structural order and all multipoles are time-even.\\
\indent Bragg spots from T $\&$ T scattering are forbidden by extinction rules. Intensities are weak compared to allowed intensities, because they are created only by electron states that possess angular anisotropy. By its very nature, T $\&$ T scattering is tailor-made for investigations of orbital ordering.\cite{ref:27}
\indent Absorption that proceeds by electric dipole transitions, $E1-E1$, reveals parity-even multipoles. In the case of an Fe ion, enhancements obtained by tuning the primary X-ray energy to L-edges expose the $3d$ ground-state $(2p \rightarrow  3d)$. Selection rules from crystal symmetry may forbid $E1-E1$, but allow weaker events, e.g., parity-odd $E1-E2$. Absorption using $E2-E2$ at the Fe K-edge also gives direct information on the $3d$ ground-state. We give explicit results for unit-cell structure factors using $E1-E1$ and $E1-E2$ events. Structure factors for an $E2-E2$ event are readily derived using expressions in the literature\cite{ref:30} and information we provide.\\
\indent Let $\left \langle O^K_Q \right \rangle$ be a Hermitian spherical multipole, with rank $K$ and projection $Q$ constrained by the condition $ -K \leq  Q \leq  K$. Angular brackets $\left \langle ... \right \rangle$ denote an expectation value, or time-average, of the enclosed tensor operator, and multipoles are properties of the ground-state of electrons. The complex conjugate of a multipole is derived from $\left \langle O^K_Q \right \rangle^*=(-1)^Q \left \langle O^K_{-Q} \right \rangle$. In Cartesian coordinates $(x, y, z)$, a rotation through an angle $\varphi $ about the $z$-axis results in the change $\left \langle O^K_Q \right \rangle\rightarrow \exp (i\varphi Q) \left \langle O^K_Q \right \rangle$. Rotations through $180^{\circ }$ about the $x$-axis and the $y$-axis result in $C_2[1, 0, 0] \left \langle O^K_Q \right \rangle \equiv  C_{2x} \left \langle O^K_Q \right \rangle = (-1)^K\left \langle O^K_{-Q} \right \rangle$ and $C_2[0, 1, 0] \left \langle O^K_Q \right \rangle \equiv  C_{2y} \left \langle O^K_Q \right \rangle = (-1)^{K + Q} \left \langle O^K_{-Q} \right \rangle$. In addition, we use identities $C_2[1, 1, 0] = C_{2y} C_{4z} = C_{4z} C_{2x}$ and $C_2[1, -1, 0] = C_{2x} C_{4z} = C_{4z} C_{2y}$.\\
\subsection{Orbital ordering has the $M^+_3$ symmetry}
\begin{figure}[t]
\includegraphics[scale=0.66]{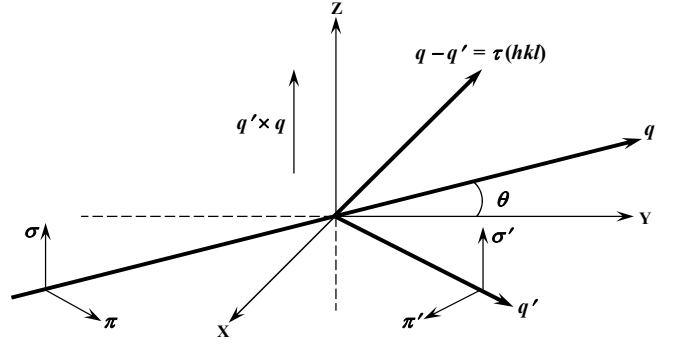}
\caption{The plane of scattering spanned by primary $(\bm {q})$ and secondary $(\bm {q'})$, and the Bragg wavevector $\bm {\tau }(hkl) = \bm {q} - \bm {q'}$. Polarization labelled $\sigma$ and $\sigma '$ is normal to the plane, and polarization labelled $\pi $ and $\pi '$ lies in the plane of scattering. The beam is deflected through an angle $2\theta$. In the nominal setting of the crystal, the $b$-axis and $c$-axis are parallel with $\bm {q} + \bm {q'}$ and $\sigma $-polarization, respectively.}
\label{fig:F7}
\end{figure}

The space group is $P4/mnc1'$ and Fe use $4d$ sites with the point group $2.221'$.\\
(i) Point group; $\left \langle O^K_Q \right \rangle$ is unchanged by $C_{2z}$, $C_2[1, 1, 0]$ and $C_2[1, -1, 0]$. We find $Q = \pm  2p$, and the identity $\left \langle O^K_{-Q} \right \rangle = (-1)^{K+p} \left \langle O^K_Q \right \rangle$. It follows that $K$ is even for $p = 0$. A monopole, $\left \langle O^0 \right \rangle$, is allowed while a dipole is forbidden, $\left \langle O^1 \right \rangle = 0$.\\
(ii) Space group; Fe sites, $(0, 1/2, 1/4)$, $(1/2, 0, 1/4)$, $(0, 1/2, 3/4)$, $(1/2, 0, 3/4)$.\\
We assign the first site with multipoles $\left \langle O^K_Q \right \rangle$ to be the reference site. Environments at the remaining three sites are generated from the reference by operations, $C_{2y}, I C_2[1, 1, 0]$ and $I C_{2y}$, respectively, in which $I$ denotes inversion. The basis of all our calculations is an electronic structure factor,
\begin{align}
\Psi^K_Q=\sum_{\bm {d}} \exp (i{\bm d}\cdot {\boldsymbol \tau }) \left \langle O^K_Q \right \rangle _{\bm {d}},
\label{eq:E2}
\end{align}
where the sum is over Fe ions at sites $\bm {d}$ in the unit cell, and the Bragg wavevector $\bm {\tau }(hkl) = (h, k, l)$ with integer Miller indices. In the result,
\begin{align}
&\Psi^K_Q(P4/mnc1')=\left \langle O^K_Q \right \rangle \exp (i \pi l /2)(-1)^K \nonumber \\
&\left [1+\sigma_{\pi }(-1)^l\right ]\left [ 1+(-1)^{h+k}(-1)^p \right ],
\label{eq:E3}
\end{align}
the parity signature of $\left \langle O^K_Q \right \rangle$ is $\sigma_{\pi }=\pm 1$. We stress that, $\Psi^K_Q(P4/mnc1')$ embodies all symmetry present in the space group, and it can be used to calculate unit-cell structure factors for nuclear scattering of neutrons, and Thomson and T $\&$ T scattering of x-rays. The same remarks apply to structure factors $\Psi^K_Q(I4/mmm1')$ and $\Psi^K_Q(I4221')$.\\
(I) Space-group allowed reflections obey $\Psi^K_Q \neq 0$ for $Q = 2p = 0$, and $\sigma_{\pi} = +1$. Extinction rules for Fe ions in space group $P4/mnc1'$ are found to be $h + k$ even and $l$ even.\\
(II) Space-group forbidden reflections $(0, 0, l)$ with $l$ odd. The electronic structure factor (\ref{eq:E3}) is different from zero for $\sigma_{\pi } = -1$. Corresponding multipoles are parity-odd and time-even, which are here denoted by $\left \langle U^K_Q \right \rangle$ and usually referred to as polar. They are visible in an $E1-E2$ event that possesses multipoles with rank $K = 1, 2, 3$.\\ 
\indent With $h = k = 0$ in (\ref{eq:E3}) the integer $p$ is even. For an $E1-E2$ event only $p = 0$ is allowed, and $K = 2$. Unit-cell structure factors, $F$, are obtained from Scagnoli and Lovesey.\cite{ref:30} They are expressed in terms of two quantities $A^K_Q = A^K_{-Q}$ and $B^K_{Q} = -B^K_{-Q}$, created from $(\Psi^K_Q + \Psi^K_{-Q})$ and $(\Psi^K_Q - \Psi^K_{-Q})$, respectively, after aligning the crystal with respect to states of polarization in the primary X-ray beam depicted in Figure \ref{fig:F7}. We find $B^2_Q = 0$ and the non-zero $A^2_Q$ are,
\begin{align}
A^2_0=-2\left \langle U^2_0 \right \rangle \exp (i\pi l /2), A^2_2=-\sqrt{3/2}A^2_0.
\label{eq:E4}
\end{align}
Note that the quadrupole $\left \langle U^2_0 \right \rangle$  is purely real.\\	
\indent Rotation of the crystal about the Bragg wavevector $(0, 0, l)$ is denoted by the (azimuthal) angle $\psi $. Unit-cell structure factors for unrotated polarization are zero, $F_{\sigma ' \sigma } = F_{\pi ' \pi } = 0$, and in rotated channels $F_{\pi ' \sigma } = -F_{\sigma ' \pi }$ is independent of the azimuthal angle, namely,
\begin{align}
F_{\pi ' \sigma }(E1-E2)=i (2/ \sqrt{5}) \cos^2 \theta \left \langle U^2_0 \right \rangle \exp (i\pi l /2),
\label{eq:E5}
\end{align}
where $\theta$ is the Bragg angle shown in Figure \ref{fig:F7}. The structure factor (\ref{eq:E5}) for T $\&$ T scattering is purely real for $l$ odd.\\
(III) 	Space-group forbidden reflections $(h, k, 0)$ with $h + k$ odd. In this case, the structure factor (\ref{eq:E3}) can be different from zero for $p$ odd, and $\sigma_{\pi }  = +1$.  We consider an $E1-E1$ absorption event. This event engages parity-even quadrupoles $\left \langle T^2_Q \right \rangle$ with $Q = \pm  2$, and we write $\left \langle T^2_{+2} \right \rangle = i\left \langle T^2_{+2} \right \rangle ''$. In Cartesian coordinates, $\left \langle T^2_{+2} \right \rangle ''$ is a quadrupole of $(xy)$-type.\\
\indent Let the Bragg wavevector $(h, k, 0)$ subtend an angle $\beta_o$ with the $a$-axis, with $\cos \beta_o = [1 + (k/h)^2]^{-1/2}$. We find the non-zero $A^2_Q$ and $B^2_Q$ are,
\begin{align}
&A^2_2=4(-1)^k  \left \langle T^2_{+2} \right \rangle '' \sin (2\beta_o), \nonumber \\ 
&B^2_2=-i4(-1)^k \left \langle T^2_{+2} \right \rangle '' \cos (2\beta_o).
\label{eq:E6}
\end{align}
Unit-cell structure factors for T $\&$ T scattering are purely real and take the values,
\begin{align}
&F_{\sigma '\sigma }(E1-E1)=- \sin^2 (\psi )A^2_2, \nonumber \\ 
&F_{\pi '\sigma }(E1-E1)=-(1/2)\sin \theta \sin(2\psi )A^2_2 + \nonumber \\
&i \cos \theta \sin(\psi ) B^2_2, \nonumber \\ 
&F_{\pi '\pi }(E1-E1)=[1-\sin^2 \theta \sin^2 (\psi )]A^2_2,
\label{eq:E7}
\end{align}
and $F_{\sigma ' \pi} (\theta ) = F_{\pi ' \sigma} (-\theta )$. In (\ref{eq:E7}) the $c$-axis is normal to the plane of scattering for $\psi  = 0$. There are no signals in unrotated channels of polarization for Bragg spots $(h, 0, 0)$ and $(0, k, 0)$ at which $\beta_o = 0$ and $\pi /2$, respectively. In the general case, structure factors for unrotated polarization are functions of $\cos (2\psi )$, whereas intensity in rotated channels has a more interesting dependence on $\psi $ because $F_{\pi ' \sigma }$ and $F_{\sigma '\pi }$ are functions of $\sin (\psi )$ and $\sin (2\psi )$. Because all unit-cell structure factors are purely real, the corresponding intensity is independent of circular polarization in the primary X-ray beam.\\
(IV) Space-group forbidden reflections $(h, 0, l)$ with $h$ and $l$ odd integers. In this case, the structure factor (\ref{eq:E3}) can be different from zero for $p$ odd, and $\sigma_{\pi }  = -1$.  There is only one value $p = 1 (Q = \pm 2)$ for an $E1-E2$ event where an octupole $(K = 3)$ is the maximum rank. T $\&$ T scattering is generated by $\left \langle U^2_{\pm 2} \right \rangle$ (purely imaginary) and $\left \langle U^3_{\pm 2} \right \rangle$ (purely real).\\
\indent $A^K_Q$ and $B^K_Q$ depend on the orientation of the Bragg wavevector with respect to crystal axes. Let $(h, 0, l)$ subtend an
 angle $\beta $ with the crystal $a$-axis with $\cos \beta  = [1 + (la/hc)^2]^{-1/2}$. Using $Z^K = 4 \left \langle U^K_{+ 2} \right \rangle \exp (i\pi l/2)$ we arrive at,
\begin{align}
&A^2_1 = -Z^2 \sin \beta , B^2_2 = Z^2 \cos \beta \nonumber \\
&A^3_0 = (\sqrt{30/2}) Z^3 \sin^2 \beta \cos \beta , \nonumber \\
&A^3_2 = (1/2) Z^3 \cos\beta (3 \cos^2 \beta - 1), \nonumber \\
&B^3_1 = -(\sqrt{10/4}) Z^3 \sin \beta (3\cos^2 \beta - 1), \nonumber \\
&B^3_3 = (\sqrt{6/4}) Z^3 \sin \beta (\cos^2 \beta + 1).
\label{eq:E8}
\end{align}
In terms of these quantities, the four unit-cell structure factors for the Bragg spot $(h, 0, l)$ with $h$ and $l$ odd are,
\begin{align}
&F_{\sigma '\sigma }(E1-E2) = (2/\sqrt{30}) \sin \theta \cos (2\psi ) A^2_1 \nonumber \\ 
&- i(1/5\sqrt{6}) \sin \theta  [5 \cos (2\psi ) + 3] B^3_1 \nonumber \\
&+i(2/\sqrt{10}) \sin \theta \sin^2 (\psi )B^3_3, \nonumber \\
&F_{\pi '\sigma }(E1-E2) = (1/2\sqrt{30}) (5 \cos 2\theta  + 1) \sin (2\psi ) A^2_1 \nonumber \\
&- (2/\sqrt{30}) \sin2 \theta \cos (\psi ) B^2_2 \nonumber \\
&+ i \sin 2\theta  \cos (\psi ) [-(1/5\sqrt{2}) A^3_0 + (1/\sqrt{15}) A^3_2] \nonumber \\ 
&+ i \sin^2 \theta  \sin (2\psi ) [(1/\sqrt{6}) B^3_1 + (1/\sqrt{10}) B^3_3], \nonumber \\
&F_{\pi '\pi }(E1-E2) = -(2/\sqrt{30}) \sin3 \theta \cos (2\psi ) A^2_1 \nonumber \\ 
&+ i(2/5\sqrt{6}) \sin \theta  [\sin^2 \theta (5 \sin^2 (\psi ) - 1) + 3 \cos^2 \theta ] B^3_1 \nonumber \\
&- i(2/\sqrt{10}) \sin \theta  [\cos^2 (\psi ) + \cos^2 \theta   \sin^2 (\psi )] B^3_3,
\label{eq:E9}
\end{align}
and $F_{\sigma ' \pi} (\theta ) = - F_{\pi ' \sigma} (-\theta )$. Note that all unit-cell structure factors are purely real, which means that the corresponding intensity is independent of circular polarization in the primary x-ray beam. The crystal $b$-axis is in the plane of scattering for $\psi  = 0$. Structure factors for unrotated polarization are functions of $\cos (2\psi )$, whereas intensity in rotated channels has a more interesting dependence on $\psi $ because $F_{\pi '\sigma}$ and $F_{\sigma '\pi }$ are functions of $\cos (2\psi )$ and $\sin (2\psi )$.\\ 
\indent A simple calculation shows that, octupoles do not contribute to $F_{\sigma '\pi }$ and $F_{\pi '\sigma }$ when $\cos^2 \beta  = 2/3$. Also, the combination of $B^3_1$ and $B^3_3$ in both $F_{\sigma '\sigma }$ and $F_{\pi '\pi }$ is independent of the azimuthal angle for the same condition on $\beta $. Using cell lengths $a = 3.91904(4)\AA$ and $c = 13.0242(3)\AA$ we find that $(l/h) = 2.35$ satisfies $\cos^2 \beta  = 2/3$. Thus, data gathered in the rotated channel for the Bragg spot $(3, 0, 7)$ can be interpreted in terms of quadrupoles alone, to a good approximation, which can then be used to extract good values for octupoles from data gathered in unrotated channels of polarization. At the Fe K-edge (7.112 keV) the Bragg spot $(3, 0, 7)$ corresponds to $sin\theta  = 0.816$.\\
\subsection{No orbital ordering}
The space group is $I4/mmm1'$ and Fe use  $4d$ sites with the point group $\bar{4}m21'$.\\
(i) Point group; $\left \langle O^K_Q \right \rangle$ is unchanged by $C_{2z}$, $I C_{4z}$ and $I C_{2x}$. We find $Q = \pm 2p, (-1)^p \sigma_{\pi } = +1$, and the identity $\left \langle O^K_{-Q} \right \rangle = (-1)^{K + p} \left \langle O^K_Q \right \rangle$.\\
(ii) The electronic structure factor is,
\begin{align}
&\Psi^K_Q(I4/mmm1') = \left \langle O^K_Q \right \rangle \exp (i\pi l/2) \nonumber \\
&\times (-1)^k [1 + (- 1)^{h + k} (-1)^p] [1 + (-1)^{h + k + l}].
\label{eq:E10}
\end{align}
Space groups $P4/mnc1'$ and $I4/mmm1'$ have the same rules for allowed reflections. And unit-cell structure factors for $(h, 0, l)$ with $h$ and $l$ odd, which are controlled by polar multipoles (\ref{eq:E9}), are the same for the two space-groups. But $(h, k, 0)$ with $h+k$ odd and $(0, 0, l)$ with $l$ odd is forbidden in $I4/mmm1'$ and allowed in $P4/mnc1'$, for which the unit-cell structure factors are given in (II).\\
\subsection{Orbital ordering has $\Gamma^-_1$ symmetry} 
The space group is $I4221'$ and Fe use  $4d$ sites with the point group $2.221'$.\\
$\Psi^K_Q(I4221') = \Psi^K_Q(I4/mmm1')$, where the latter is given in (\ref{eq:E10}), and both $(h, k, 0)$ with $h+k$ odd  and $(0, 0, l)$ with $l$ odd are forbidden. A distinguishing feature of $I4221'$ is that both parity-even and parity-odd events can contribute to the Bragg spot $(h, 0, l)$ with $h$ and $l$ odd.\\ 
\indent Consider an $E1-E1$ event and define $Z^2 = i4 \left \langle T^2_{+2} \right \rangle '' \exp (i\pi l/2)$, which is purely real for $l$ odd. Unit-cell structure factors are written in terms of $A^2_1 = -Z^2 \sin \beta $ and $B^2_2 = Z^2 \cos \beta$, where $\beta $ is the angle subtended by $(h, 0, l)$ and the $a$-axis. We find,
\begin{align}
&F_{\sigma '\sigma }(E1-E1) =  -i \sin (2\psi ) A^2_1, \nonumber \\
&F_{\pi '\sigma } (E1-E1)= -i \sin \theta  \cos (2\psi ) A^2_1 + \nonumber \\
&i \cos \beta  \sin (\psi ) B^2_2, \nonumber \\
&F_{\pi '\pi }(E1-E1) = -i \sin^2 \theta  \sin (2\psi ) A^2_1 
\label{eq:E11}
\end{align}
and $F_{\sigma '\pi }(\theta ) = F_{\pi '\sigma} (-\theta )$. Notice that the dependence of structure factors on the azimuthal angle is different for $E1-E2$ and $E1-E1$ events at $(h, 0, l)$ with $h$ and $l$ odd; comparing (\ref{eq:E9}) for $E1-E2$ and (\ref{eq:E11}) for $E1-E1$ we see that $\cos (2\psi ) \Leftrightarrow  \sin (2\psi )$ and $\cos (\psi ) \Leftrightarrow  \sin (\psi )$.
\section{Conclusion}
\indent Structural properties of iron-based superconductors have been discussed, using the symmetry methods formulated with the Landau theory of phase transitions. Two mechanisms, namely, magnetic and orbital ordering, for symmetry lowering in the orthorhombic and the newly-discovered re-entrant tetragonal phases are considered in detail. The key result of the present study is the identification of distinct space group symmetries for the re-entrant tetragonal phase, predicted by magnetic and orbital ordering mechanisms. This provides a direct way to experimentally reveal the  underlying physical mechanism through a precise structural determination available at modern diffraction facilities.\\
\indent The magnetic mechanism with in-plane magnetic moments implies magneto-elastic coupling resulting in the atomic displacements and orbital ordering which reduce the crystallographic space group symmetry (space group without magnetic subsystem) down to $P4/mmm1'$. The symmetry lowering can be detected by conventional diffraction methods through an observation of $h+k+l$ odd reflections with $l \neq 0$. The magnetic mechanism with out-of-plane magnetic moments, as found in Ba$_{0.76}$Na$_{0.24}$Fe$_2$As$_2$ from the present neutron powder diffraction experiment, implies a two-$\bm{k}$ magnetic structure which does not allow any orbital ordering and the crystal structure symmetry (without magnetic subsystem) of the system is well approximated by the parent $I4/mmm1'$ space group. An experimental confirmation of the two-$\bm{k}$ nature of the magnetic structure (for instance in a single crystal neutron diffraction experiment with uniaxial strain) would provide strong evidence for the magnetic scenario and the relevance of the itinerant electronic model.\\
\indent The orbital ordering mechanism does not require the magnetic structure to be two-$\bm{k}$ and predicts the crystal structure symmetry lowering down to $P4/mnc1'$ or $ I4221'$ depending on the stacking of the $(ab)$ ordered layers along the $c$-axis. Both types of orbital ordering do not allow any atomic displacements in comparison with the parent $I4/mmm1'$ space group but all three space groups can be distinguished in X-ray resonant scattering by inspecting the $(h,k,0)$ with $h+k$ odd, $(0,0,l)$ with $l$ odd and $(h,0,l)$ with $h$ and $l$ odd reflections, in respect of the presence of T $\&$ T scattering and the parity of the multipoles contributing to the diffraction. The first two families of reflections are expected to be non-zero only in the case of the $P4/mnc1'$ symmetry. The third type of the reflections can distinguish the $I4/mmm1'$ and $I4221'$ space groups.  The orbital patters with the $P4/mnc1'$ and $ I4221'$ symmetries cannot be induced by the magnetic order parameter and can appear only as independent instability. An observation of these patterns in X-ray resonant scattering would provide strong evidence for the orbitally-driven scenario.

\indent Work by  O.C., R.O, and S.R. was supported by the Materials Science and Engineering Division, Basic Energy Sciences, Office of Science, U.S. Department of Energy.

\end{document}